\documentclass[iop,apj]{emulateapj}
\usepackage{amsmath}
\usepackage[varg]{txfonts}
\usepackage{braket}
\usepackage{bm}

\usepackage[OT2,OT1]{fontenc}
\newcommand\cyr{
\renewcommand\rmdefault{wncyr}
\renewcommand\sfdefault{wncyss}
\renewcommand\encodingdefault{OT2}
\normalfont
\selectfont}
\DeclareTextFontCommand{\textcyr}{\cyr}

\newtheorem{theo}{Theorem}[section]
\newtheorem{lemma}[theo]{Lemma}
\newtheorem{corol}[theo]{Corollary}

\newcommand\Nu{\mathrm N}
\renewcommand\today{May 16, 2011}

\begin{document}

\journalinfo{accepted for the publication in
\it the Astrophysical Journal \rm (\today)}

\title{Constraints on velocity anisotropy of
spherical systems\\with separable augmented densities}
\shorttitle{\sc Separable Augmented Densities\hfill}
\author{J.~An\altaffilmark{1}}\shortauthors{\hfill\sc J.~An}
\altaffiltext{1}
{National Astronomical Observatories, Chinese Academy of Sciences,
A20 Datun Rd., Chaoyang Dist., Beijing 100012, P.R.China;}
\slugcomment{\tt jinan@nao.cas.cn}

\begin{abstract}\noindent
If the augmented density of a spherical anisotropic system is
assumed to be multiplicatively separable to functions of the
potential and the radius, the radial function, which can be
completely specified by the behavior of the anisotropy parameter
alone, also fixes the anisotropic ratios of every higher-order
velocity moment. It is inferred from this that the non-negativity
of the distribution function necessarily limits the allowed
behaviors of the radial function. This restriction is translated
into the constraints on the behavior of the anisotropy parameter.
We find that not all radial variations of the anisotropy
parameter satisfy these constraints and thus that there exist
anisotropy profiles that cannot be consistent with any
separable augmented density.
\end{abstract}
\keywords{galaxies: kinematics and dynamics ---
methods: analytical --- dark matter}

\section{INTRODUCTION}

One basic problem in stellar dynamics is to find the distribution
function that is consistent with the given local density profile.
For the simplest cases, the problem reduces to solving an
integral equation. The solution to finding the ergodic
distribution function for an isotropic spherical system, i.e.,
the so-called \citeauthor{Ed16} formula is known to
as early as its namesake.
On the other hand, the complete solutions to devising the
two-integral even distribution function for an axisymmetric
system are also available through the works
by \cite{Fr52}, \citet{Ly62}, and \citet{HQ93} etc.

By contrast, the construction of a two-integral
distribution function for an anisotropic spherical system contains
an additional difficulty. This is because integrating the two-integral
distribution function over the velocity space produces an
\emph{augmented density}, which is a bivariate function of the potential
and the radius. While it is an easy exercise to
demonstrate that deducing the distribution function from the
augmented density is a formally identical problem to the case of
the axisymmetric
system with a two-integral even distribution function \citep[see
e.g.,][]{Qi93,An11}, it is also obvious from the onset that,
given the local density and the potential, there is no unique way
to specify the bivariate augmented density without any additional
prescription regarding the system. In fact, if the potential is
known, specifying the augmented density is essentially equivalent
to knowing an infinite subset of the velocity moment
functions \citep[cf.,][]{DM92}, to the zeroth of which the local
density corresponds. The usual line of attack is to restrict
either the augmented density or the distribution function to be
in the specified functional form
\citep[e.g.,][]{Os79,Me85,Cu91,CP95,AE06a,Wo08} and match further
properties such as the anisotropy parameter to that following the
particular ansatz. One advantage of this approach is that these
procedures usually simplify the subsequent inversion for the
distribution function from the augmented density although these
tend to sacrifice the flexibility in the behaviors of the varying
anisotropy.

On the other hand, the procedure that uniquely specifies the
system and also allows the greatest possible freedom for the
radially varying anisotropy has been outlined in \citet{QH95}.
If the potential- and radial-dependences of the augmented density
are assumed to be multiplicatively separable, then the radial
part can be determined by the anisotropy parameter alone. The
potential part then follows immediately once the potential and
the density are specified. In fact, demanding the separable
augmented density is the only route that fixes the anisotropy
parameter, independently of the knowledge on the potential and
the density. A practical implementation utilizing a general
parametric form of the monotonically varying anisotropy parameter
is found in \citet{BvH07}, although adopting any parametric
form intrinsically restricts the full flexibility of the technique.

Recently, the separable augmented density has been attracted
renewed interests in the context of the necessary conditions for
the distribution function to be non-negative. For instance, \citet{CM10}
proposed, for any system with a separable augmented density,
the existence of the so-called global density slope--anisotropy
inequality, which conditionally extends the central density
slope--anisotropy theorem of \citet{AE06} to all radii.
This has been subsequently proved with a restriction on the central
anisotropy by \citet{vHBD} and \citet{An11}.

This paper further explores the implications of the separability
in the augmented density and its limitation. In particular, we
examine the relation between the distribution function and the
moment functions and derive how the anisotropy parameter together
with the potential and density uniquely specifies the separable
augmented density. We also find that the radial part of the
separable augmented density completely determines the anisotropic
behaviors of not only the velocity dispersions but also every
higher-order even velocity moment. Based on these findings, we
also show that the non-negativity of the moment functions, which
follows the non-negativity of the distribution function, restricts
the physically permitted behaviors of the separable augmented
density and the anisotropy parameter described by it.
In the following discussion, we consider our constraints in the context
of the necessary and sufficient conditions for a separable augmented
density corresponding to a physical system, extending prior works.
Finally, further discussion concerning the inversion for
the anisotropy profile in relation to the separable augmented density
and our constraints is also provided.

\section{PRELIMINARY}

The \citeauthor{Je15} theorem indicates that
a steady-state spherical dynamical system is described by
the phase-space distribution function (DF) of the form of
$f(\mathcal E,L^2)$. Here $\mathcal E$, the specific binding
energy, and $L$, the magnitude of the specific angular momentum,
are the two classical isotropic isolating integrals admitted
by the spherical potential, i.e.,
$\mathcal E=\Psi(r)-\frac12v^2$ and
$L=(\bm{L\cdot L})^\frac12=rv_\mathrm t$
where $\Psi$ is the relative potential with respect
to the boundary (Hence, $\mathcal E>0$ for a bound particle).
Finally, $v=(v_\mathrm t^2+v_r^2)^\frac12$ with
$v_\mathrm t=(v_\theta^2+v_\phi^2)^\frac12$ and $v_r$
being the tangential and radial velocities. Note that
$(v_r,v_\theta,v_\phi)$ constitutes the set of three orthogonal
velocity components in a common unit.

Integrating the DF over the velocity space,
\begin{equation}\label{eq:aden}\begin{split}
\Nu\bigl(\Psi,r^2\bigr)&\equiv
 \iiint_{v^2\le2\Psi}\!d^3\!\bm v\,
  f\bigl(\mathcal E,L^2\bigr)
\\&=\frac{2\pi}r\!
 \iint_{\substack{\mathcal E\ge0,L^2\ge0\\
                 2r^2\mathcal E+L^2\le2r^2\Psi}}\!
 \frac{f\,d\mathcal E\,dL^2}
      {\sqrt{2r^2(\Psi-\mathcal E)-L^2}},
\end{split}\end{equation}
results in a bivariate function of $\Psi$ and $r^2$. This is
usually referred to as the ``augmented density'' (AD). Once the potential
$\Psi=\Psi(r)$ (which is not necessarily generated by the following
density) is specified, the local density is found to be
$\nu(r)=\Nu[\Psi(r),r^2]$. In a self-consistent system on the other hand,
the Poisson equation with the AD as the source term
results in an ordinary differential equation on $\Psi(r)$, which can be
solved to determine $\nu(r)$ uniquely.

The local higher-order velocity moments are also found similarly.
Whereas any odd-integral moment must vanish thanks to the
spherical symmetry (in particular, the isotropy in the configuration
space), all the even-integral moments are found to be
$\overline{v_r^{2p}v_\mathrm t^{2q}}=m_{p,q}/\Nu$ where
\begin{subequations}
\begin{equation}\begin{split}
m_{p,q}\bigl(\Psi,r^2\bigr)&\equiv
 \iiint_{v^2\le2\Psi}\!d^3\!\bm v\,
  v_r^{2p}v_\mathrm t^{2q}f\bigl(\mathcal E,L^2\bigr)
\\&=\frac{2\pi}{r^{2q+2}}\!
 \iint_\mathcal T\!d\mathcal E\,dL^2
  \mathcal K^{p-\frac12}L^{2q}f.
\end{split}\end{equation}
Here the transform kernel is given by
\begin{equation}
\mathcal K\bigl(\mathcal E,L^2;\Psi,r^2\bigr)\equiv
2(\Psi-\mathcal E)-\frac{L^2}{r^2},
\end{equation}
which is actually $v_r^2$ expressed as a function of the 4-tuple
$(\mathcal E,L^2;\Psi,r^2)$, whereas the integral is over the region
in $(\mathcal E,L^2)$ space defined to be
\begin{equation}
\mathcal T\equiv
\set{(\mathcal E,L^2)|\mathcal E\ge0,L^2\ge0,\mathcal K\ge0},
\end{equation}
\end{subequations}
that is, the triangular region bounded by lines $\mathcal E=0$,
$L^2=0$ and $\mathcal K=0$. The last line is the same as the
diagonal line given by $\mathcal E+(2r^2)^{-1}L^2=\Psi$.

If $p>\frac12$, we find that
\begin{equation}\label{eq:dif}
\frac{\partial\bigl(r^{2q+2}m_{p,q}\bigr)}{\partial X}=
(2p-1)\,\pi\!
\iint_\mathcal T\!d\mathcal E\,dL^2\,
\mathcal K^{p-\frac32}\frac{\partial\mathcal K}{\partial X}L^{2q}f
\end{equation}
where $X=\Psi$ or $r^2$. Given that
$\frac{\partial\mathcal K}{\partial\Psi}=2$ and
$\frac{\partial\mathcal K}{\partial(r^2)}=\frac{L^2}{r^4}$,
equation~(\ref{eq:dif}) indicates the existence of differential
recursion relations for the moment functions 
\begin{subequations}\label{eq:rr}\begin{align}
&\frac{\partial m_{p,q}}{\partial\Psi}=(2p-1)m_{p-1,q};
\label{eq:rr1}\\\label{eq:rr2}
&\frac{\partial\bigl(r^{2q+2}m_{p,q}\bigr)}{\partial r^2}
=\bigl(p-\tfrac12\bigr)\,r^{2q}m_{p-1,q+1},
\end{align}\end{subequations}
which is valid for $p>\frac12$. In fact, once the AD
is specified, every other velocity moment can be recovered without
inverting for the DF. Specifically, given
$m_{0,0}=\Nu$ and the `initial conditions'
$m_{p,0}(0,r^2)=0$, we first find that, for $k\ge1$
\begin{subequations}
\begin{equation}\begin{split}
m_{k,0}(\Psi,r^2)&=2^k\bigl(\tfrac12\bigr)_k\overbrace{
\int_0^\Psi\!d\Psi_k\,\dotsi\!\int_0^{\Psi_2}\!d\Psi_1}^\text{$k$ times}
m_{0,0}(\Psi_1,r^2)
\\&=\frac{2^k\bigl(\frac12\bigr)_k}{(k-1)!}\!
\int_0^\Psi\!dQ\,(\Psi-Q)^{k-1}\Nu(Q,r^2),
\end{split}\end{equation}
via repeated integrations of equation~(\ref{eq:rr1}) and
the Cauchy formula for repeated integration (eq.~\ref{a6}).
Here $(a)_n=\prod_{i=1}^n(a+i-1)$
is the \emph{rising} sequential product (the Pochhammer symbol).
Next, the repetitions of equation (\ref{eq:rr2}) yield
\begin{equation}
\biggl[\prod_{j=1}^q\bigl(k+\tfrac12-j\bigr)\biggr]\,m_{k-q,q}
=\frac1{r^{2q+2}}\,
\biggl(r^4\!\frac\partial{\partial r^2}\biggr)^q\bigl(r^2m_{k,0}\bigr),
\end{equation}
for $0\le q\le k$.
Combining these and using equation~(\ref{th3}), we recover
the result of \citet[eq.~13]{DM92},
\begin{multline}\label{eq:p}
m_{k-q,q}(\Psi,r^2)
\\=\frac{2^k\bigl(\frac12\bigr)_{k-q}}{(k-1)!}\!
\int_0^\Psi\!dQ\,(\Psi-Q)^{k-1}
\biggl(\frac\partial{\partial r^2}\biggr)^q\bigl[r^{2q}\Nu(Q,r^2)\bigr]
\end{multline}
\end{subequations}
for $k\ge1$ and $0\le q\le k$. The proper verification that
equation~(\ref{eq:p}) is the unique solution to equations~(\ref{eq:rr})
given $m_{p,q}(0,r^2)=0$ is provided in Appendix~\ref{app:A}.
An immediate corollary following equation~(\ref{eq:p}) is that
\begin{equation}
\biggl(\frac\partial{\partial r^2}\biggr)^q
\bigl[r^{2q}\Nu(\Psi,r^2)\bigr]\ge0
\end{equation}
for every non-negative integer $q$ is a sufficient (but not necessary)
condition for every $m_{p,q}$ to be non-negative.

The behavior of the anisotropic velocity dispersions in a spherical
system is usually parametrized by the ``velocity anisotropy parameter''
\citep{Bi80},
\begin{subequations}
\begin{equation}\label{eq:beta}
\beta(r)\equiv1-\frac{\sigma_\mathrm t^2}{2\sigma_\mathrm r^2}
=1-\frac{m_{0,1}[\Psi(r),r^2]}{2m_{1,0}[\Psi(r),r^2]}
\end{equation}
where $\sigma_\mathrm r^2=\overline{v_r^2}$ and
$\sigma_\mathrm t^2=\overline{v_\mathrm t^2}
=\overline{v_\theta^2+v_\phi^2}$.
Meanwhile, equation~(\ref{eq:rr2}) with $(p,q)=(1,0)$ reduces to
\begin{equation}
\frac{\partial\bigl(r^2m_{1,0}\bigr)}{\partial r^2}=\frac{m_{0,1}}2.
\end{equation}
Hence, the anisotropy parameter is directly related to the radial
partial derivative of the moment function $m_{1,0}$, i.e.,
\begin{equation}\label{eq:betam}
\beta=1-\frac{m_{0,1}}{2m_{1,0}}
=-\frac{\partial\ln m_{1,0}}{\partial\ln r^2}\biggr\rvert_{\Psi(r),r^2}.
\end{equation}
\end{subequations}
Then, the total radial derivative of $m_{1,0}$ results in
\begin{equation}\label{eq:je0}\begin{split}
\frac{dm_{1,0}}{dr}
&=\frac{\partial m_{1,0}}{\partial r}
+\frac{\partial m_{1,0}}{\partial\Psi}\frac{d\Psi}{dr}
\\&=\frac{2m_{1,0}}r\frac{\partial\ln m_{1,0}}{\partial\ln r^2}
+m_{0,0}\frac{d\Psi}{dr}.
\end{split}\end{equation}
With $\Psi=\Psi(r)$ and equation~(\ref{eq:betam}),
equation~(\ref{eq:je0}) is simply the second-order steady-state
spherical Jeans equation. Note in fact that $\Psi(r)=\Phi_0-\Phi(r)$
is the \emph{relative} potential with respect to the reference
$\Phi_0$ where $\Phi(r)$ is the true gravitational potential,
and thus $-\frac{d\Psi}{dr}=\frac{d\Phi}{dr}=\frac{GM_r}{r^2}$ where
$M_r$ is the enclosed gravitating mass within the sphere of radius $r$.

More generally, for $p>\frac12$, equation~(\ref{eq:rr2}) indicates that
\begin{equation}
\frac{\partial m_{p,q}}{\partial r}
=-\frac2r\,\Bigl[(q+1)m_{p,q}-\bigl(p-\tfrac12\bigr)\,m_{p-1,q+1}\Bigr].
\end{equation}
The corresponding total radial derivatives result in
\begin{multline}\label{eq:je}
\frac{dm_{p,q}}{dr}
=-\frac2r\,\Bigl[(q+1)m_{p,q}-\bigl(p-\tfrac12\bigr)\,m_{p-1,q+1}\Bigr]
\\+(2p-1)m_{p-1,q}\frac{d\Psi}{dr},
\end{multline}
which in fact constitute the complete set of the Jeans equations
\citep{DM92} -- see also \citet{MK90} for the fourth-order equations,
which correspond to $(p,q)=(2,0)$ and $(1,1)$ here.

\section{SEPARABLE AUGMENTED DENSITY}

Let us suppose that the $\Psi$- and $r^2$-dependences of
the AD are multiplicatively separable as in
\begin{equation}
\Nu(\Psi,r^2)=P(\Psi)R(r^2),
\end{equation}
for some $P(\Psi)$ and $R(r^2)$. It then follows from
equation~(\ref{eq:p}) that every moment function $m_{p,q}(\Psi,r^2)$
is also separable. In particular,
$m_{p,q}(\Psi,r^2)=2^{p+q}(\tfrac12)_pP_{p+q}(\Psi)R_q(r^2)$ or
\begin{subequations}
\begin{equation}
m_{k-n,n}(\Psi,r^2)=2^k\bigl(\tfrac12\bigr)_{k-n}P_k(\Psi)R_n(r^2)
\end{equation}
for $0\le n\le k$, where
\begin{align}
P_k(\Psi)&\equiv\begin{cases}
P(\Psi)&(k=0)\\
\displaystyle{\frac1{(k-1)!}\!\int_0^\Psi\!dQ\,(\Psi-Q)^{k-1}P(Q)}
&(k\ge1)\end{cases}
\label{eq:funp}\\\label{eq:funr}
R_n(r^2)&\equiv
\biggl(\frac d{dr^2}\biggr)^n\bigl[r^{2n}R(r^2)\bigr]
=\frac1{r^{2n+2}}\,
\biggl(r^4\!\frac d{dr^2}\biggr)^n\bigl[r^2R(r^2)\bigr].
\end{align}
\end{subequations}
We find from equation~(\ref{eq:funp}) that
$\frac d{d\Psi}P_k=P_{k-1}$ for any positive integer $k$. Similarly,
equations~(\ref{eq:funr}) and (\ref{th3}) lead to
$\frac d{d(r^2)}(r^{2n+2}R_n)=r^{2n}R_{n+1}$
for any non-negative integer $n$.
Next, since $m_{1,0}(\Psi,r^2)=P_1(\Psi)R(r^2)$,
equation~(\ref{eq:betam}) indicates
\begin{subequations}\begin{align}
&\beta(r)=-\frac{d\ln R(r^2)}{d\ln r^2}\,;
\label{eq:sbeta}\\\label{eq:sbeta2}
&\frac{R(r^2)}{R(\hat r^2)}=
\exp\biggl\lgroup
2\!\int_r^{\hat r}\!\frac{\beta(\tilde r)}{\tilde r}
\,d\tilde r\biggr\rgroup.
\end{align}\end{subequations}
In other words, the radial function $R(r^2)$ is completely specified
(up to an immaterial scale constant) given the anisotropy parameter
$\beta(r)$. Once $R$ is specified, the potential part
immediately follows the local density as
$P(\Psi)=\nu(r)/R(r^2)$ with the inverse function $r=\Psi^{-1}(\Psi)$
of the potential \citep{QH95,BvH07}.

\subsection{Implications of the separable augmented density}
\label{sec:imp}

Given the boundary conditions $\Psi(r_0)=0$ and $m_{1,0}(0,r_0^2)=0$
at $r=r_0$ (which may be the infinity), the radial velocity dispersion
is given by
\begin{subequations}\begin{align}
\nu\sigma_\mathrm r^2\bigr|_r
&=m_{1,0}\bigl[\Psi(r),r^2\bigr]=R(r^2)\int_0^{\Psi(r)}\!dQ\,P(Q)
\\&=R(r^2)\int_{r_0}^r\!\frac{\nu(\hat r)}{R(\hat r^2)}
\frac{d\Psi}{dr}\biggr\rvert_{\hat r}d\hat r.
\label{eq:pres}\end{align}\end{subequations}
Here equation~(\ref{eq:pres}) is actually the solution to
the steady-state spherical Jeans equation with $R^{-1}$
of equation~(\ref{eq:sbeta2}) being its integrating factor
\citep[e.g.,][]{vdM94,AE09}. In other words, equation~(\ref{eq:pres})
always provides the velocity dispersions of the system given
the potential, the density, and the anisotropy parameter
irrespective of the separability assumption.

However, the true implications of the separability assumption
on the other hand lie beyond the behaviors of the velocity dispersions.
That is to say, with the separable AD assumption,
the anisotropy parameter not only specifies the complete AD
together with the local density and the potential, but also it
constrains the anisotropic behaviors of \emph{every} higher-order
velocity moment (including naturally those of velocity dispersions)
by itself. In particular,
\begin{subequations}\label{eq:nbeta0}
\begin{equation}\label{eq:nbeta}
\alpha_n\equiv\frac{R_n}{R_0}
=\bigl(\tfrac12+p\bigr)_n\,\frac{m_{p,n}}{m_{p+n,0}},
\end{equation}
while $\alpha_n(r)$ for any non-negative integer $n$ is determined
recursively from $\beta(r)$ alone (Appendix~\ref{app:B}) via
\begin{equation}\label{eq:alp}
\alpha_{n+1}=(n+1-\beta)\alpha_n+\alpha^\prime_n
\end{equation}
with $\alpha_0=1$. Here,
$\alpha_n^\prime=\frac{d\alpha_n}{d\ln r^2}=
\frac r2\frac{d\alpha_n}{dr}$.
For a few small $n$'s, this is specifically translated into
\begin{align}
\alpha_1
&=1-\beta=\frac{m_{0,1}}{2m_{1,0}}=\frac{3m_{1,1}}{2m_{2,0}}
=\frac{5m_{2,1}}{2m_{3,0}}
=\dotsb;
\\
\alpha_2
&=(1-\beta)(2-\beta)-\beta^\prime
=\frac{3m_{0,2}}{4m_{2,0}}=\frac{15m_{1,2}}{4m_{3,0}}
=\dotsb;
\\
\alpha_3
&=(1-\beta)(2-\beta)(3-\beta)
-3(2-\beta)\beta^\prime-\beta^{\prime\prime}
=\frac{15m_{0,3}}{8m_{3,0}}
=\dotsb,
\end{align}\end{subequations}
and so on. Here, 
$\beta^{\prime\prime}=\frac{d^2\beta}{du^2}=\frac{d\beta^\prime}{du}$
where $u=\ln r^2$ etc.

Furthermore, we also have
\begin{subequations}\begin{align}
\frac{dm_{p,q}}{dr}=&\
\frac d{dr}\biggl(\frac{m_{p,q}}{m_{p+q,0}}m_{p+q,0}\biggr)
\nonumber\\&
=\frac{C_{p,q}R_q}{C_{p+q,0}R_0}\frac{dm_{p+q,0}}{dr}
+C_{p,q}P_{p+q}R_0\frac{d\alpha_q}{dr};
\\
\frac{\partial m_{p,q}}{\partial r}=&\
C_{p,q}P_{p+q}\frac d{dr}\biggl(\frac{R_q}{R_0}R_0\biggr)
\nonumber\\&
=C_{p,q}P_{p+q}\frac{R_q}{R_0}\frac{dR_0}{dr}
+C_{p,q}P_{p+q}R_0\frac{d\alpha_q}{dr};
\\
\frac{\partial m_{p,q}}{\partial\Psi}=&\
C_{p,q}\frac{dP_{p+q}}{d\Psi}R_q
=C_{p,q}P_{p+q-1}R_q
\end{align}\end{subequations}
where $C_{p,q}=2^{p+q}(\tfrac12)_p$. Therefore, expressing
the total radial derivative of $m_{p,q}(\Psi,r^2)$ leads to
(note $R_0=R$)
\begin{subequations}
\begin{equation}
\frac{dm_{p+q,0}}{dr}=C_{p+q,0}\,\biggl[P_{p+q}\frac{dR}{dr}
+P_{p+q-1}R\frac{d\Psi}{dr}\biggr].
\end{equation}
Given equation~(\ref{eq:sbeta}), this indicates that,
if the AD is assumed to be separable, the set
of the ($2n$)-th order spherical Jeans equations (eq.~\ref{eq:je}
with $p+q=n$ and $p\ge1$) reduces to a single equation,
\begin{equation}\label{eq:s4JE}
\frac{dm_{n,0}}{dr}+\frac{2\beta}rm_{n,0}
=(2n-1)m_{n-1,0}\frac{d\Psi}{dr}.
\end{equation}
\end{subequations}
This generalizes the fourth-order Jeans equation for `constant
anisotropy' introduced by \citet[see also \citealt{LM03}]{Lo02}.
We note however that equation~(\ref{eq:s4JE}) is in fact
the result of the separability assumption and not of the
constant anisotropy per se. In addition, under the separability
assumption, the solution to equation~(\ref{eq:s4JE}) is immediately
obvious as per $m_{n,0}=(2n-1)!!P_n[\Psi(r)]R(r^2)$ with
equations~(\ref{eq:funp}) and (\ref{eq:sbeta2}) as well as
$P[\Psi(r)]=\nu(r)/R(r^2)$.

\setcounter{footnote}0
\subsection{Constraints on the anisotropy parameter}
\subsubsection{general cases with a separable augmented density}

The non-negativity of the DF implies that all
the even-integral moment functions must be also non-negative.
Consequently, if $\Nu(\Psi,r^2)=P(\Psi)R(r^2)$ is separable,
then $P_n(\Psi)\ge0$ and $R_n(r^2)\ge0$ for any non-negative
integer $n$. While $P(\Psi)\ge0$ is the sufficient (and also the
necessary since $P=P_0$) condition for $P_n(\Psi)\ge0$ for any
non-negative integer $n$, the condition that
\begin{subequations}\label{eq:main}\begin{gather}
R_n(r^2)=
\frac{d^n\bigl[x^nR(x)\bigr]}{dx^n}\biggr\rvert_{x=r^2}\ge0
\quad\text{($r,x\ge0$)}
\intertext{or equivalently (see eq.~\ref{th3})}
\biggl(r^4\!\frac d{dr^2}\biggr)^n\bigl[r^2R(r^2)\bigr]
=(-1)^n\frac{d^n}{dw^n}
\frac{R(w^{-1})}w\biggr\rvert_{w=r^{-2}}\ge0
\quad\text{($r,w\ge0$)}
\end{gather}\end{subequations}
for every non-negative integer $n$ constitutes a set of
independent constraints on the behavior of $R(r^2)$.\footnote{A
function $\phi(x)$ of $x>0$ is said to
be ``completely monotonic'' (c.m.) if and only if $(-1)^n\phi^{(n)}(x)\ge0$
for all non-negative integers $n$. Hence, the condition
is equivalent to saying $\mathcal R(w)=R(w^{-1})/w$ is a c.m.\
function of $w$. According to S.~Bernstein's theorem
on c.m.\ functions \citep[see][]{Wi41},
an important corollary to this is that $\mathcal R(w)$ must be
the Laplace transformation of a non-negative function -- i.e.,
the inverse Laplace transformation of $\mathcal R(w)$ exists and
is non-negative for positive reals. Moreover,
expressing the inverse Laplace transformation using
E.~Post's inversion formula \citep[see][]{HW55}
indicates that eq.~(\ref{eq:main}) is actually equivalent to
$\lim_{n\rightarrow\infty}R_n\bigl(\frac tn\bigr)/n!\ge0$ for $t\ge0$.
Further explorations of this idea will be given elsewhere.}
In other words, equation~(\ref{eq:main}) is a necessary
condition for the radial part $R(r^2)$ of any separable AD
to be generated by a non-negative DF.

Combined with equation~(\ref{eq:sbeta}), it forms the set of
restrictions on the radial variations of $\beta(r)$
allowed for the spherical system with a separable AD.
In particular, any spherical anisotropic system with a separable
AD is physical only if
\begin{subequations}\label{eq:cons}
\begin{equation}
\alpha_n(r)\ge0
\quad\text{($r\ge0$)}
\end{equation}
for all positive integers $n$. Here the set of functions $\alpha_n$
is as defined in equation~(\ref{eq:alp}) with
$\beta(r)$. For the first few small $n$'s, the conditions are equivalent to
\begin{align}
&\beta\le1;\label{eq:cons1}
\\&\frac r2\frac{d\beta}{dr}\le(1-\beta)(2-\beta);\label{eq:cons2}
\\&\frac r4\frac d{dr}
\Bigl(r\frac{d\beta}{dr}\Bigr)
+\frac32(2-\beta)\,r\frac{d\beta}{dr}
\le(1-\beta)(2-\beta)(3-\beta),
\end{align}\end{subequations}
and so on.
Equation~(\ref{eq:cons1}) is obvious from the definition of the anisotropy
parameter (eq.~\ref{eq:beta}) and the non-negativity of the velocity
dispersions, and thus universal independently of the separability
assumption.
By contrast, the further constraints involving the radial derivatives of
$\beta$ are the consequence of the separability assumption
-- following equation~(\ref{eq:nbeta0}) and the non-negativity of the
higher-order velocity moments.

These imply that, even if $R(r^2)$ could
be formally written down using equation~(\ref{eq:sbeta2}), not all
arbitrarily varying $\beta(r)$ are consistent with separable
AD because some might produce negative higher-order
moments. For example, consider the anisotropy parameter behaving
\begin{subequations}
\begin{equation}\label{eq:omb0}
\beta(r)=\frac{r^{2s}}{r_\mathrm a^{2s}+r^{2s}}
\end{equation}
so that the system is isotropic at the center and radially
anisotropic in the outskirts. If $s=1$, the anisotropy profile
of equation~(\ref{eq:omb0}) is that of the
\citeauthor{Os79}-\citeauthor{Me85} (OM) system.
The corresponding radial function is
$R(r^2)=(1+r^{2s}/r_\mathrm a^{2s})^{-1/s}$ within a constant, but
this does not satisfy the condition in equation~(\ref{eq:main}) if $s>1$
because 
\begin{equation}
\frac{d^2\bigl[x^2R(x)\bigr]}{dx^2}\biggr\rvert_{x=r^2/r_\mathrm a^2}
=\frac{2-(s-1)x^s}{(1+x^s)^{\frac1s+2}}
\end{equation}
which is negative for $x>\sqrt[s]{2/(s-1)}$.
Equivalently we find
\begin{equation}
(1-\beta)(2-\beta)-\frac r2\frac{d\beta}{dr}
=\frac{2r_\mathrm a^{2s}-(s-1)r^{2s}}{(r_\mathrm a^{2s}+r^{2s})^2},
\end{equation}
\end{subequations}
and thus equation (\ref{eq:omb0}) fails the constraint in
equation~(\ref{eq:cons2}) if $s>1$ and $r/r_\mathrm a>\sqrt[2s]{2/(s-1)}$.
Consequently, $\beta(r)$ in equation~(\ref{eq:omb0})
is consistent with a separable AD only if $s\le 1$.
In fact, the converse is also true, i.e., if $s\le 1$,
then $R(r^2)=(1+r^{2s}/r_\mathrm a^{2s})^{-1/s}$ satisfies
equation~(\ref{eq:main}) for all non-negative integers $n$ -- obviously,
if $s=1$, the non-negative OM DF
exists with a properly chosen potential term $P(\Psi)$.

Roughly, equation~(\ref{eq:cons2}) insists that the anisotropy
parameter in the system with a separable AD cannot increase
radially faster than the limiting value determined by the local
anisotropy parameter, which tends to get smaller as it becomes more radially
anisotropic. Similar interpretations for higher-order constraints
of equation~(\ref{eq:cons}) are less obvious.

\subsubsection{a family of monotonic anisotropy parameters}

Consider the anisotropy profile,
\begin{subequations}\label{eq:pbeta0}
\begin{equation}\label{eq:pbeta}
\beta(r)=
\frac{\beta_\infty r^{2s}+\beta_0r_\mathrm a^{2s}}
     {r^{2s}+r_\mathrm a^{2s}}
\quad\text{($s>0$)}.
\end{equation}
This parametrization has also been introduced by \citet{BvH07} for their
construction of dynamical models with a flexible anisotropy parameter.
Equation~(\ref{eq:omb0}) corresponds to equation~(\ref{eq:pbeta})
with $(\beta_0,\beta_\infty)=(0,1)$. Note that the transform
of $s\rightarrow-s$ is actually equivalent to switching
$\beta_0\leftrightarrow\beta_\infty$, and thus the restriction $s>0$
is actually not necessary. Nevertheless, to assign definite physical
meanings to the parameters, we retain the restriction. Then $\beta$
monotonically varies from $\beta_0$ at the center to $\beta_\infty$
as $r\rightarrow\infty$, with the constant-$\beta$ case represented by
$\beta_0=\beta_\infty$. The choice that $r_\mathrm a=0$ or
$r_\mathrm a=\infty$ also produces the constant-$\beta$ model although
they will not be considered explicitly here. The case $s=\frac12$ reduces
to the generalized \citeauthor{ML05} anisotropy model \citep[cf.,][]{MB10}
with the original \citet{ML05} model given by
$(\beta_0,\beta_\infty)=(0,\frac12)$. The separable AD
with $s=1$ include the \citeauthor{Cu91} system \citep[see also][]{CM10a}
for which $\beta_\infty=1$ and the OM
system with $(\beta_0,\beta_\infty)=(0,1)$. As \citet{BvH07}
have noticed, this parametrization is notable as it yields the simple
analytic integrating factor for the second-order Jeans equation
(eq.~\ref{eq:sbeta2}),
\begin{equation}\label{eq:bres}
[R(r^2)]^{-1}=
r^{2\beta_0}(r^{2s}+r_\mathrm a^{2s})^{\frac{\beta_\infty-\beta_0}s}.
\end{equation}
\end{subequations}
With $\nu(r)$ and $\Psi(r)$ specified, the radial velocity
dispersion can be found in quadrature by equation~(\ref{eq:pres}),
regardless of the separability of the AD.

Under the separable AD assumption on the other hand,
this completely specifies the resulting system;
$\Nu(\Psi,r^2)=P(\Psi)R(r^2)$ where
$P[\Psi(r)]=\nu(r)/R(r^2)$ and $R(r^2)$ is deduced from
equation~(\ref{eq:bres}).
The DF can be found by inverting the integral
equation~(\ref{eq:aden}) -- e.g., \citet{De86} and \citet{BvH07} for
the technique based on the Laplace-Mellin transform or
\citet{Qi93} and \citet[see also \citealt{An11}]{HQ93}
for the complex contour integral method.

However, the preceding arguments indicate that the resulting model is
not necessarily physical for an arbitrary parameter set. Obviously, the
condition that $\beta(r)\le1$ for $\forall r\ge0$ restricts the parameters
to be $\beta_0\le1$ and $\beta_\infty\le1$.
As the system with a separable AD,
more constraints on the parameters
also follow equations~(\ref{eq:main}) and (\ref{eq:cons}).
The first of these corresponding to equation~(\ref{eq:main}) with $n=2$
reduces to
\begin{subequations}\label{eq:conb0}
\begin{multline}
\bigl[(2-\beta_0)+(2-\beta_\infty)x^s\bigr]\,
\bigl[(1-\beta_0)+(1-\beta_\infty)x^s\bigr]
\\\ge s\,(\beta_\infty-\beta_0)\,x^s
\quad\text{(for $x\ge0$)},
\end{multline}
which is also equivalent to equation (\ref{eq:cons2});
\begin{multline}
(r^{2s}+1)^2\biggl[(1-\beta)(2-\beta)-\frac r2\frac{d\beta}{dr}\biggr]
\\=\bigl[(2-\beta_\infty)r^{2s}+(2-\beta_0)\bigr]\,
\bigl[(1-\beta_\infty)r^{2s}+(1-\beta_0)\bigr]
\\-s\,(\beta_\infty-\beta_0)\,r^{2s}\ge0
\quad\text{(for $r\ge0$)}.
\end{multline}
Here, we have set $r_\mathrm a=1$ for brevity, but this does not affect
the following results.
Since the necessary and sufficient condition for the
real-coefficient monic quadratic equation $x^2+bx+c=0$ to possess
no non-degenerate positive real root is $c\ge0$ and $b\ge-2|c|^{\frac12}$,
equation (\ref{eq:conb0}) for $\beta_0,\beta_\infty\le1$
is also equivalent to
\begin{multline}
(2-\beta_\infty)(1-\beta_0)+(2-\beta_0)(1-\beta_\infty)
-s(\beta_\infty-\beta_0)\\+2(2-\beta_0)^\frac12(1-\beta_0)^\frac12
(2-\beta_\infty)^\frac12(1-\beta_\infty)^\frac12\ge0.
\end{multline}
\end{subequations}
With
$(2-\beta_\infty)(1-\beta_0)+(2-\beta_0)(1-\beta_\infty)
=2(2-\beta_0)(1-\beta_\infty)+\beta_\infty-\beta_0$,
we therefore find for fixed $\beta_0,\beta_\infty\le1$ that
equation~(\ref{eq:conb0}) \emph{fails} if $\beta_0<\beta_\infty$ \emph{and}
\begin{multline}
s>1+
\frac{2(2-\beta_0)^\frac12(1-\beta_\infty)^\frac12}
     {\beta_\infty-\beta_0}
\\\times
\Bigl[(2-\beta_0)^\frac12(1-\beta_\infty)^\frac12
+(1-\beta_0)^\frac12(2-\beta_\infty)^\frac12\Bigr].
\end{multline}
That is, there exist parameter combinations for
equation~(\ref{eq:pbeta0}) that cannot be consistent with any
physical separable AD.

More higher-order constraints may be derived similarly,
but direct calculations for general cases become rather
complicated as the order increases. Instead, here we just
note that, if $0<s\le 1$ or $\frac{\beta_0-\beta_\infty}s$ is a
non-negative integer, then $R(r^2)$
in equation~(\ref{eq:bres}) satisfies the condition of
equation~(\ref{eq:main}) and so $\beta(r)$
in equation~(\ref{eq:pbeta}) can be consistent with a physical separable
AD. An elementary proof is provided in Appendix~\ref{app:C}.
The sufficiency of the condition that $0<s\le1$ for the
parametrization given in equation~(\ref{eq:pbeta0}) can also be deduced by
the existence of the corresponding non-negative DF
with a separable AD as demonstrated by \citet{BvH07},
who explicitly constructed the particular DF
in terms of the convergent Fox H-function. 
We also suspect that if $\beta_0<\beta_\infty$, the
condition that $0<s\le 1$ is the necessary condition
for equation~(\ref{eq:bres}) to satisfy equation~(\ref{eq:main})
but have no definite proof at this time.

In addition, we also note that the condition in equation~(\ref{eq:main}) is
linear on $R(r^2)$. Hence, if both $A(r^2)$ and $B(r^2)$ meet the
necessary condition in equation (\ref{eq:main}), the radial function given
by the linear combination,
$R(r^2)=aA(r^2)+bB(r^2)$ where $a$ and $b$ are positive
constants, also satisfies the same necessary condition (and therefore the
anisotropy parameter resulting from it is consistent with eq.~\ref{eq:cons}
and a physical separable AD). For example, this
indicates that the \emph{multicomponent} generalized
\citeauthor{Cu91} systems studied by
\citet{CM10a} do satisfy equation~(\ref{eq:main}) as their radial
functions are given by the sums of the functions in the form of
equation~(\ref{eq:bres}) with $s=1$ (and $\beta_\infty=1$) and different
$r_\mathrm a$'s.

\section{DISCUSSION}
\subsection{Sufficient conditions for physical separable
augmented densities}

For multicomponent Cuddeford systems
\citet{CM10a} have proved that the condition
$\frac{d^{\mu+1}P}{d\Psi^{\mu+1}}\ge0$ is sufficient to guarantee
the non-negativity of the posited DF. Here $\mu$
is the integer floor of (i.e., the greatest integer not larger than)
$\frac32-\beta_0$. Subsequently, \citet{vHBD} asked whether the same
condition should be the sufficient condition for any separable AD
to be generated by the non-negative DF.

The present paper clarifies the answer to their question in the simplest
form to be negative. The existence of the necessary condition involving
only the radial function $R(r^2)$, i.e., equation~(\ref{eq:main}), implies
that any sufficient condition must also contain some restrictions on the
same. The hypothesis as stated entirely with the potential part $P(\Psi)$
thus cannot be a sufficient condition by itself given the independent
nature of the potential and radial parts of the separable AD.
A simple counterexample may be constructed with equation~(\ref{eq:pbeta})
and the choice of parameters such that $\frac12<\beta_0<\beta_\infty=1$
and $s>1$. With the radial function $R(r^2)$ that follows
(eq.~\ref{eq:bres}), no function $P(\Psi)$, regardless of
$\frac{dP}{d\Psi}\ge0$ or not,
can lead to a non-negative DF -- here the choice of
$P(\Psi)$ is equivalent to specifying the local density as
$P[\Psi(r)]=\nu(r)/R(r^2)$ and insisting $\frac{dP}{d\Psi}\ge0$ imposes
the so-called global density slope anisotropy inequality \citep{CM10,vHBD}.
Nonetheless, the findings in this paper do not preclude the possibility
that the constraint $\frac{d^{\mu+1}P}{d\Psi^{\mu+1}}\ge0$ combined with
additional conditions on $R(r^2)$ may constitute a sufficient condition
for the system with a separable AD. In fact, after the
original version of this paper was submitted, E. Van Hese (private
communication) has discovered the existence of the set of such conditions,
e.g., together $\frac{dP}{d\Psi}>0$ and the set of conditions on $R(r^2)$
that includes equation~(\ref{eq:main}) are sufficient for
the existence of a non-negative DF.

\subsection{Universal constraints on the anisotropy parameter?}

Since the constraints in equation~(\ref{eq:cons}) are actually put
on the anisotropy parameter without any explicit reference to the separable
AD, it seems fair to ponder how general these constraints
actually are -- i.e., whether the constraints exist for any physical
AD. Although we find, for the time being,
no reason to argue that these
constraints are universal (for they are derived based on the particular
assumption of the separable AD) beyond the obvious
restriction that $\beta\le1$, we will not attempt to settle the answer
in this paper. However, we do note that, if the constraints
are entirely the consequence of the separability of the AD,
one must be able to construct a pair of the non-negative DF
and the \emph{inseparable} AD for a spherical
anisotropic system with its anisotropy parameter violating
the conditions in equation~(\ref{eq:cons}).
Unfortunately, the task is complicated by the fact that,
with inseparable AD, the anisotropy parameter cannot be
specified independently without imposing the fixed behavior of the
potential; the pair of $f(\mathcal E,L^2)$ and $\Nu(\Psi,r^2)$ typically
prescribes only $\beta(\Psi,r^2)$ and thus varying $\Psi=\Psi(r)$
results in a different $\beta(r)=\beta[\Psi(r),r^2]$unless
$\Psi(r)$ and $\nu(r)=\Nu[\Psi(r),r^2]$ are related to
each other through the Poisson equation and so the freedom to choose
$\Psi(r)$ is subsequently removed.

\subsection{Separable augmented densities and the Jeans degeneracy}

With real data, our observations are typically limited by projection,
and thus the usual kinematical observables available to us are restricted
to the surface brightness profile (or the column density profile for
the discrete number count data) and the line-of-sight (los) velocity
dispersion. It is a well-known fact that, while the three-dimensional
density profile can be uniquely inverted from the surface density
(the Abel transformation) under the spherical symmetry assumption, the
radial and tangential velocity dispersions, $\sigma_\mathrm r^2(r)$ and
$\sigma_\mathrm t^2(r)$ -- which are needed to find the potential through
the Jeans equation -- cannot be determined from the los velocity
dispersion alone as they are degenerate in reproducing the observations of
the last \citep{De87,Me87} unless the system is known to have
isotropic velocity distributions.
One may lift this so-called Jeans degeneracy
by imposing additional constraints on the system coming from observations
or a priori assumptions.

For example, \citet{MB10} and \citet{Wo10} have shown that it is in general
possible to find $\sigma_\mathrm r^2$ that is consistent with the observed
los velocity dispersion profile and any arbitrarily specified
anisotropy parameter $\beta(r)$. \citet{EAW} on the other hand demonstrated
that one can also go in the other way around, finding $\beta$ that is
consistent with the observed los velocity dispersion and the
arbitrary assumed form of $\sigma_\mathrm r^2$.
Alternatively, under the assumption of the constant mass-to-light ratio,
one can also find the unique solution to the coupled Jeans-Poisson equations
from the los velocity dispersion profile 
\citep{BM82,To83,Bi89}. This is equivalent to specifying
the potential first and inverting the los velocity dispersion
to determine $\beta$ given the Jeans equation \citep{SS90,DM92}.
However, for the purpose of constraining the
potential, these anisotropy inversion algorithms can only,
at best, reject some choices for the gravitational potential as unphysical
where $\beta$ reaches values above unity (formally this indicates negative
velocity dispersion).

The best observational constraint for lifting the Jeans degeneracy
on the other hand would be some handle on the proper motions of
tracers \citep[e.g.,][]{LM89,vdMA10}
as they are the velocity projections that are orthogonal to the
los velocity. Alternatively, with a data set consisting of
discrete tracers, the precisely measured differential distances (which
ultimately yield the distances to the center of the system) can also
break the Jeans degeneracy \citep[cf.,][]{WEA}. Unfortunately, with our
current and near-future observational capabilities, their uses are mostly
limited to very near-by objects.

A popular idea for possible observational constraints is the use of the
higher-order moments \citep{MK90} or the distribution of the
los velocities -- note that specifying the distribution is
equivalent to knowing the infinite set of the entire moments. Similar to
the velocity dispersions (which are the second moments),
\citet{DM92} have shown that, with the potential specified,
the complete set of independent velocity moments can be solved from
the observed los velocity moments up to the same order
-- in the infinite order, this implies
that the DF is uniquely specified by the observed
distribution of the los velocities provided that the potential is
known a priori. However, it is easy to argue that this will not solve the
degeneracy problem (in particular for tracing the potential 
observationally)
because, under the spherical symmetry, introducing each new
($2n$)-th moment adds ($n+1$) new variables and the $n$ constraining Jeans
equations
(eq.~\ref{eq:je}) with one further observational constraint into the
mix and therefore there is no net increase in the constraints.

\citet{Lo02} and \citet{LM03} on the other hand introduced a hybrid of
theoretical and observational constraints, `constant anisotropy' and the
fourth moments (kurtosis), to bring the degeneracy problem into the
unique solution. However, their `constant anisotropy' is actually in the
form of a strictly stronger assumption that the DF and
the AD are given by the ansatz\footnote{Recently,
\citet{Wo08} proposed an extension of this by introducing a more
general ansatz for the $L$ part of the DF that allows
the variation of $\beta$, which they found to be
consistent with the DF of simulated $\Lambda$CDM halos.
In a sequel, \citet{Wo09} predicted the
distribution of the los velocities with which the observed data
can be fit to determine the parameters of the DF.
In principle, this is still less flexible than the procedure outlined
in the following, but if in practice one were to \emph{parametrize}
the anisotropy parameter in a particular functional form, the approach
may be seen as complementary.};
$f(\mathcal E,L^2)=L^{-2\beta}g(\mathcal E)$ and
$\Nu(\Psi,r^2)=r^{-2\beta}P(\Psi)$
while \S~\ref{sec:imp} of the present paper
(in particular, eq.~\ref{eq:s4JE}) indicates that this is a rather
unnecessarily restrictive assumption for their method to work. That is to
say, under the separability assumption of the AD, to bring
higher-than-the-second-order moments into the problem only adds one
independent new variable (cf., eq.~\ref{eq:nbeta}) and the single Jeans
equation (eq.~\ref{eq:s4JE}). Therefore the observations
of the los velocity moment at the same order
actually act as an additional net
constraint on the system given the separable AD.
Specifically, the introduction of the fourth moment is enough to
uniquely solve the Jeans degeneracy if the AD is assumed to
be separable whereas adding further higher-order moments actually
over-constrains the problem.

However, the assumption of a separable AD
is purely formal and its physical interpretation is unclear,
although it is a weaker hypothesis than the power-law ansatz for
the $L$ part of DF and the AD
which produces the constant anisotropy. The most conservative
statement that can be drawn regarding the Jeans degeneracy and
the separable
AD is thus that given the observations of the second and
fourth moments (the dispersion and the kurtosis) of the
los velocities, there exists a unique spherical model with a
separable AD that is consistent with them. The resulting
model is complete in that it essentially specifies the DF
as well as the underlying potential. Although the non-negativity of the
resulting DF is not guaranteed, the condition
in equation~(\ref{eq:main}) is both necessary and sufficient
to prove that the model
will produce non-negative velocity moments of every order -- of course,
the model is not necessarily `real' and the predicted
higher-than-fourth-order moments should be compared to the observations
(if available) in order for it to be acceptable.

\small
\acknowledgments\noindent
The author thanks
Maarten Baes, Steen Hansen, Lucia Morganti, and
Emmanuel Van Hese for their comments on the earlier version
and discussion of the subject in general.
The author also appreciates the very detailed report
by the anonymous referee whose suggestions have
helped to improve the present paper.
The author is supported by the Chinese Academy of Sciences (CAS)
Fellowships for Young International Scientists (Grant No.:2009Y2AJ7), and
the National Natural Science Foundation of China (NSFC) Research Fund for
International Young Scientists.

\medskip
\section*{appendix}
\setcounter{section}0
\numberwithin{equation}{section}
\renewcommand\thesection{\Alph{section}}
\renewcommand\theequation{\thesection\arabic{equation}}
\section{Proof of equation (\lowercase{\ref{eq:p}})}
\label{app:A}

\begin{lemma}\label{th1}
For a non-negative integer $n$ and arbitrary real $a$,
\begin{equation}
\frac{d^n(x^a)}{dx^n}=\biggl[\prod_{j=0}^{n-1}(a-j)\biggr]\,x^{a-n},
\end{equation}\end{lemma}
which is easily proved by the induction on $n$. For a positive
integer power monomial (i.e., $k$ is a non-negative integer),
this simplifies
\addtocounter{equation}{-1}\begin{subequations}\begin{equation}
\frac{d^n(x^k)}{dx^n}=\begin{cases}\
\displaystyle{\frac{k!x^{k-n}}{(k-n)!}}&(0\le n\le k)\smallskip\\\
0&(n\ge k+1)\end{cases}.
\end{equation}\end{subequations}
Next, using Lemma~\ref{th1} and the extended Leibniz rule,
we find that
\begin{lemma}
for a non-negative integer $n$ and any function $f$,
\begin{equation}
\frac{d^n\bigl(x^nf\bigr)}{dx^n}
=\sum_{k=0}^n\binom nk\,
\frac{d^{n-k}(x^n)}{dx^{n-k}}\frac{d^kf}{dx^k}
=\sum_{k=0}^n\frac{(n!)^2}{(k!)^2(n-k)!}x^kf^{(k)}.
\end{equation}\end{lemma}
Here $\binom nk$ is the binomial coefficient and
$f^{(k)}(x)=\frac{d^kf}{dx^k}$.
Now, we are able to prove
\begin{theo}
for a non-negative integer $n$ and any function $f$,
\begin{equation}\label{th3}
\biggl(x^2\!\frac d{dx}\biggr)^n\bigl(xf\bigr)
=x^{n+1}\frac{d^n\bigl(x^nf\bigr)}{dx^n}.
\end{equation}
\end{theo}
{\it Proof.} We prove this by the induction on $n$.
First, equation (\ref{th3}) is trivial for $n=0,1$.
The induction step is proved as
\addtocounter{equation}{-1}
\begin{subequations}\begin{equation}\begin{split}
\biggl(x^2&\!\frac d{dx}\biggr)^{n+1}\bigl(xf\bigr)
=x^2\frac d{dx}\Biggl[\biggl(x^2\!\frac d{dx}\biggr)^n\bigl(xf\bigr)\Biggr]
=x^2\frac d{dx}\biggl[x^{n+1}\frac{d^n\bigl(x^nf\bigr)}{dx^n}\biggr]
\\&=x^2\frac d{dx}
\biggl[\sum_{k=0}^n\frac{(n!)^2}{(k!)^2(n-k)!}x^{n+1+k}f^{(k)}\biggr]
\\&=x^2\sum_{k=0}^n
\frac{(n!)^2\bigl[(n+1+k)x^{n+k}f^{(k)}+x^{n+k+1}f^{(k+1)}\bigr]}
     {(k!)^2(n-k)!}
\\&=x^2\sum_{k=0}^{n+1}
\frac{(n!)^2\bigl[(n+1-k)(n+1+k)+k^2\bigr]}{(k!)^2(n-k+1)!}x^{n+k}f^{(k)}
\\&=x^{n+2}\sum_{k=0}^{n+1}\frac{(n!)^2(n+1)^2}{(k!)^2(n-k+1)!}x^kf^{(k)}
=x^{n+2}\frac{d^{n+1}\bigl(x^{n+1}f\bigr)}{dx^{n+1}}.
\end{split}\end{equation}\end{subequations}
Immediately following this is
\begin{corol}
for any non-negative integer $k$,
\begin{equation}\label{co4}\begin{split}
\frac d{dx}\biggl[x^{k+1}\frac{d^k\bigl(x^kf\bigr)}{dx^k}\biggr]
&=\frac d{dx}\Biggl[\biggl(x^2\!\frac d{dx}\biggr)^k\bigl(xf\bigr)\Biggr]
\\&=\frac1{x^2}\biggl(x^2\!\frac d{dx}\biggr)^{k+1}\bigl(xf\bigr)
=x^k\frac{d^{k+1}\bigl(x^{k+1}f\bigr)}{dx^{k+1}}.
\end{split}\end{equation}\end{corol}
With $f=\Nu(\Psi,r^2)$, $x=r^2$, and $k=q$, this results in
\addtocounter{equation}{-1}
\begin{subequations}\begin{equation}\label{th4}
\frac\partial{\partial r^2}\biggl[r^{2q+2}
\Bigl(\frac\partial{\partial r^2}\Bigr)^q
\bigl[r^{2q}\Nu(\Psi,r^2)\bigr]\biggr]
=r^{2q}\Bigl(\frac\partial{\partial r^2}\Bigr)^{q+1}
\bigl[r^{2q+2}\Nu(\Psi,r^2)\bigr].
\end{equation}\end{subequations}
It is now easy to show that $m_{p,q}(\Psi,r^2)$ in
equation~(\ref{eq:p}) satisfies equation~(\ref{eq:rr2}) by direct
calculations using equation~(\ref{th4}) and the Pochhammer symbol
$(\frac12)_p=(\frac12)_{p-1}(p-\frac12)$ for any positive integer $p$.

One can show that equation~(\ref{eq:p}) satisfies
equation~(\ref{eq:rr1}) using
\begin{subequations}\begin{equation}
\frac d{dx}\!\int_{x_0}^x\!dy\,(x-y)^kg(y)
=\delta_{k,0}\,g(x)+k\!\int_{x_0}^x\!dy\,(x-y)^{k-1}g(y)
\end{equation}
(where $\delta_{m,n}$ is the Kronecker delta) and
\begin{equation}\begin{split}
\int_{x_0}^x\!dx_1\!\int_{x_0}^{x_1}\!dy\,(x_1-y)^kg(y)
&=\int_{x_0}^x\!dy\,g(y)\!\int_y^x\!dx_1\,(x_1-y)^k
\\&=\frac1{k+1}\!\int_{x_0}^x\!dy\,(x-y)^{k+1}g(y)
\end{split}\end{equation}
for any $k\ge0$. By repeatedly applying this, one finds that
a simple iterated integral in general reduces to an integral transform,
\begin{equation}\label{a6}
g_k(x)=\underbrace{
\int_0^x\!dx_k\,\dotsi\!\int_0^{x_2}\!dx_1}_\text{$k$ times}g(x_1)
=\frac1{(k-1)!}\!\int_0^x\!dy\,(x-y)^{k-1}g(y),
\end{equation}\end{subequations}
where $k$ is now a positive integer. This is sometimes known as
the Cauchy formula for repeated integration, and can be strictly
proved through the induction on $k$. In fact, the function
$g_k(x)$ defined as such is the particular solution to
the differential equation $d^kg_k(x)/dx^k=g(x)$ with
the set of initial conditions $g_k^{(j)}(0)=0$ where
$j\in\set{0,\dotsc,k-1}$.
Formally, equation~(\ref{a6}) extends to the case $k=0$ by noting that
$\lim_{\epsilon\rightarrow0^+}\epsilon|x|^{\epsilon-1}=2\delta(x)$
where $\delta(x)$ is the Dirac delta.
The extra factor of two is due to the fact
that the integral interval in equation~(\ref{a6})
is one-sided extending down from $x$.

\section{General expression for equations
(\lowercase{\ref{eq:nbeta0}})\&(\lowercase{\ref{eq:cons}})}
\label{app:B}

For any function $f(x)$ of $x$, we find using equation (\ref{co4}) that
\begin{equation}\begin{split}
\frac{d^{n+1}\bigl[x^{n+1}f(x)\bigr]}{dx^{n+1}}
&=\frac1{x^n}
\frac d{dx}\biggl[x^{n+1}\frac{d^n\bigl(x^nf\bigr)}{dx^n}\biggr]
\\&=\biggl(1+n+x\frac d{dx}\biggr)\,\frac{d^n\bigl[x^nf(x)\bigr]}{dx^n}.
\end{split}\end{equation}
%
%
Next, with the definitions of $\alpha_n$ in equation~(\ref{eq:nbeta})
and of $R_n$ in equation~(\ref{eq:funr})
\begin{equation}
\alpha_n(r)=\frac{R_n(r^2)}{R_0(r^2)}
=\frac1R\frac{d^n\bigl[x^nR(x)\bigr]}{dx^n}\biggr\rvert_{x=r^2}
\end{equation}
from the given radial function $R(r^2)$ of a separable AD,
the recursion formula for $\alpha_n$ in equation~(\ref{eq:alp}) follows as
\begin{equation}\begin{split}
\alpha_{n+1}&=\frac1R\,
\biggl(1+n+x\frac d{dx}\biggr)\,\bigl(R\alpha_n\bigr)\biggr\rvert_{x=r^2}
\\&=\biggl(1+n+\frac{d\ln R}{d\ln r^2}\biggr)\,\alpha_n
+\frac{d\alpha_n}{d\ln r^2}.
\end{split}\end{equation}
Replacing $R$ with $\beta$ utilizing equation~(\ref{eq:sbeta})
yields equation~(\ref{eq:alp}).
Equation~(\ref{eq:main}) then indicates equation~(\ref{eq:cons})
whereas equation~(\ref{eq:alp}) may be
considered to be the recursive definition of $\alpha_n(r)$ from
the anisotropy parameter $\beta(r)$ given the initial term $\alpha_0=1$.
Note also that $\alpha_n$ in equation~(\ref{eq:alp}) is defined as such
without referring to the radial function or the separable AD at all.

\section{Proof of Consistency of equation (\lowercase{\ref{eq:pbeta0}})}
\label{app:C}

For $\beta(r)$ and $R(x)$ given by equations~(\ref{eq:pbeta0}),
if we define
\begin{equation}
\tau_n\equiv(1+x^s)^n\alpha_n
=x^{\beta_0}(1+x^s)^{n+\lambda}\frac{d^n}{dx^n}
\biggl[\frac{x^{n-\beta_0}}{(1+x^s)^\lambda}\biggr],
\end{equation}
where $\lambda=\frac{\beta_\infty-\beta_0}s$,
equation~(\ref{eq:alp}) results in
\begin{equation}\label{eq:trec}
\tau_{n+1}
=\bigl[(n+1-\beta_0)+(n+1-\beta_\infty-sn)y\bigr]\tau_n
+sy(1+y)\frac{d\tau_n}{dy},
\end{equation}
where $y\equiv x^s=\frac{r^{2s}}{r_\mathrm a^{2s}}$. Since
$\tau_0=\alpha_0=1$, this indicates that $\tau_n$ is an
(at most) $n$-th order polymonial of $y$. If we then let
\begin{equation}\label{eq:poly}
\tau_n=\sum_{k=0}^n\tilde t_{n,k}y^k,
\end{equation}
we can derive the recursion relation for the coefficients,
\begin{subequations}
\begin{equation}\label{eq:tcr}
\tilde t_{n+1,k}=
(n+1-\beta_0+sk)\tilde t_{n,k}
+\bigl[(1-s)n+1-\beta_\infty+s(k-1)\bigr]\tilde t_{n,k-1}
\end{equation}
by substituting equation~(\ref{eq:poly}) into
equation~(\ref{eq:trec}), and also using $\tilde t_{n,k}=0$
for $k<0$ or $k>n$. If $k=0$, equation~(\ref{eq:tcr})
reduces to
\begin{equation}
\tilde t_{n+1,0}=(n+1-\beta_0)\tilde t_{n,0}
\qquad\Rightarrow\
\tilde t_{n,0}=(1-\beta_0)_n\ge0
\end{equation}\end{subequations}
provided that $\beta_0\le1$ because $\tilde t_{n,-1}=0$ and
$\tilde t_{0,0}=1$. Next, for a positive integer pair
$n\ge k\ge1$, equation~(\ref{eq:tcr}) indicates that,
if $0<s\le1$,
the non-negativity of $\tilde t_{n-1,k}$ and $\tilde t_{n-1,k-1}$
can guarantee the non-negativity of $\tilde t_{n,k}$ provided that
$\beta_0,\beta_\infty\le1$. Since we have already found that
$\tilde t_{n,k}=0$ for $k<0$ and $\tilde t_{n,0}=(1-\beta_0)_n\ge0$,
we can conclude
that if $0<s\le1$ and $\beta_0,\beta_\infty\le1$, then
$\tau_n(y)$ is a polynomial with all non-negative coefficients
and therefore $\tau_n(y)\ge0$ for ${}^\forall y\ge0$ and any
non-negative integer $n$.
Since $\tau_n=(1+x^s)^n\alpha_n=(1+x^s)\frac{R_n}R$,
equation~(\ref{eq:main}) also follows immediately.

As for the cases that $\xi=\frac{\beta_0-\beta_\infty}s=-\lambda$ is
a non-negative integer, we first consider the constant-$\beta$ case,
that is, $\xi=0$ and $\beta_0=\beta_\infty=\beta$. Then,
using equation~(\ref{th1}),
\begin{subequations}
\begin{equation}
R_n(x)=\frac{d^nx^{n-\beta}}{dx^n}
=\biggl[\prod_{j=0}^{n-1}(n-\beta-j)\biggr]\,x^{-\beta}
=\frac{(1-\beta)_n}{x^\beta}\ge0,
\smallskip\end{equation}
for $x>0$, provided that $\beta\le1$. In general,
if $\xi=\frac{\beta_0-\beta_\infty}s$ is a non-negative integer,
we can simply extend this result to
\begin{align}
R&=\frac{(1+x^s)^\xi}{x^{\beta_0}}
=\sum_{k=0}^\xi\binom\xi k\,x^{sk-\beta_0}
\\R_n&
=\sum_{k=0}^\xi\binom\xi k\,\frac{d^n(x^{sk+n-\beta_0})}{dx^n}
=\sum_{k=0}^\xi\binom\xi k\,(sk+1-\beta_0)_nx^{sk-\beta_0}.
\end{align}
\end{subequations}
Again, $R_n(x)\ge0$ for $x>0$, provided that
$\beta_\infty=\beta_0-s\xi\le\beta_0\le1$.

\end{document}